\documentclass[twocolumn]{aastex62}

\usepackage{apjfonts}
\usepackage{amsmath}

\newcommand{\Hi}{\hbox{H{\sc i}}}

\begin{document}

\title{The evolution of baryonic mass function of galaxies to $z=3$}
\author{Zhizheng Pan}
\email{panzz@pmo.ac.cn}
\affiliation{Purple Mountain Observatory, Chinese Academy of Sciences, 10 Yuan Hua Road, Nanjing, Jiangsu 210033, China}
\affiliation{School of Astronomy and Space Sciences, University of Science and Technology of China, Hefei, 230026, China
}

\author{Yingjie Peng}
\affiliation{Kavli Institute for Astronomy and Astrophysics, Peking University, Yi He Yuan Lu 5, Hai Dian District, Beijing 100871, China}

\author{Xianzhong Zheng}
\email{xzzheng@pmo.ac.cn}
\affiliation{Purple Mountain Observatory, Chinese Academy of Sciences, 10 Yuan Hua Road, Nanjing, Jiangsu 210033, China}
\affiliation{School of Astronomy and Space Sciences, University of Science and Technology of China, Hefei, 230026, China
}

\author{Jing Wang}
\affiliation{Kavli Institute for Astronomy and Astrophysics, Peking University, Yi He Yuan Lu 5, Hai Dian District, Beijing 100871, China}

\author{Xu Kong}
\email{xkong@ustc.edu.cn}
\affiliation{School of Astronomy and Space Sciences, University of Science and Technology of China, Hefei, 230026, China
}
\affiliation{CAS Key Laboratory for Research in Galaxies and Cosmology, Department of Astronomy, \\
University of Science and Technology of China, Hefei, Anhui 230026, China}

\begin{abstract}
We combine the published stellar mass function (SMF) and gas scaling relations to explore the baryonic (stellar plus cold gas) mass function (BMF) of galaxies to redshift $z=3$. We find evidence that at log$(M_{\rm baryon}/M_{\sun})>11.3$, the BMF evolves little since $z\sim 2.2$. With the evolution of BMF and SMF, we investigate the baryon net accretion rate ($\dot{\rho}_{\rm baryon}$) and stellar mass growth rate ($\dot{\rho}_{\rm star}$) for the galaxy population of log($M_{\rm star}/M_{\sun}$)>10. The ratio between these two quanties, $\dot{\rho}_{\rm baryon}$/$\dot{\rho}_{\rm star}$, decreases from $\dot{\rho}_{\rm baryon}$/$\dot{\rho}_{\rm star}\sim$2 at $z\sim 2.5$ to $\dot{\rho}_{\rm baryon}$/$\dot{\rho}_{\rm star}<$0.5 at $z\sim 0.5$, suggesting that massive galaxies are transforming from the ``accretion dominated" phase to the ``depletion dominated" phase from high$-z$ to low$-z$. The transition of these two phases occurs at $z\sim1.5$, which is consistent with the onset redshift of the decline of cosmic star formation rate density. This provides evidence to support the idea that the decline of cosmic star formation rate density since $z\sim1.5$ is mainly resulted from the decline of baryon net accretion rate and star formation quenching in galaxies.

\end{abstract}

\section{Introduction} \label{sec:intro}
The distribution of baryonic (stellar plus cold gas) mass of galaxies is of fundamental importance for studying the assembly of galaxies over cosmic time. The first attempt for studying the baryonic mass function (BMF) of galaxies was done by \citet{Bell 2003}, who found that the local BMF is almostly identical to the stellar mass function (SMF) at the high-mass end. This is straightforward to interpret since the baryon content of local massive galaxies has been dominated by stars. In the low-mass regime, the BMF has a similar low-end slope as the SMF. Similar features are also found by later studies that based on different galaxy samples \citep{Papastergis 2012,Eckert 2016}. To date, the investigation of BMF is limited to the local Universe.

With the advent of deep surveys in the past two decades, the investigation of SMF has now been pushed out to redshift $z=8$ \citep{Ilbert 2010,Ilbert 2013,Muzzin 2013, Tomczak 2014,Song 2016,Davidzon 2017}. In the meantime, new observations have facilitated the study of gas properties of high-redshift galaxies in more details. Generally, galaxies tend to have higher gas fraction towards higher redshifts \citep{Tacconi 2010, Tacconi 2013,Gowardhan 2019}. Specifically, \citet{Tacconi 2013} found that at $z\sim 2.2$, the ratio between gas mass and total baryonic mass, $f_{\rm gas}=M_{\rm gas}/(M_{\rm gas}+M_{\rm star})$, is around 50\% for a galaxy with log$(M_{\rm star}/M_{\sun})=11.0$. Given this, stellar mass maybe no longer dominate the baryonic budget of a galaxy even at the high mass end in the early universe.

In this paper, we aim to combine the newly published SMF of \citet{Davidzon 2017} and the gas-scaling relations of \citet{Tacconi 2018}, to push the investigation of BMF to $z=3$. In Section 2, we introduce the methodology used in this work. In Section 3, we present the derived BMF. In Section 4, we compare the baryon net accretion rate and stellar mass growth rate for galaxies with log($M_{\rm star}/M_{\sun}$)>10. A short summary and discussion are presented in Section 5.  Throughout this paper, we adopt a concordance $\Lambda$CDM cosmology with $\Omega_{\rm m}=0.3$, $\Omega_{\rm \Lambda}=0.7$, $H_{\rm 0}=70$ $\rm km~s^{-1}$ Mpc$^{-1}$ and a \citet{Chabrier 2003} initial mass function (IMF). $M_{\rm gas}$ of this work refers to the gas mass of atomic plus molecular hydrogen in the inter-stellar medium (ISM), which have included a correction of 1.36 to account for helium.

\section{Methodology}
Galaxies with a same $M_{\rm star}$ may have different $M_{\rm gas}$. Given this, galaxies of similar $M_{\rm star}$ could exhibit a broad distribution in the $M_{\rm star}+M_{\rm gas}$ (hereafter $M_{\rm baryon}$) space. For galaxies within each $M_{\rm star}$ bin, once their $M_{\rm baryon}$ distribution is determined, then at a fixed baryonic mass of $M_{\rm baryon}$, the number density of galaxies can be derived using the following equation:
\begin{equation}
\Phi(M_{\rm baryon})=\sum^{N}_{i=1}\Phi(M_{\rm baryon}|M_{\rm i})
\end{equation}
, where $\Phi({M_{\rm baryon}|M_{i}})$ is the number density of galaxies that with a baryonic mass of $M_{\rm baryon}$ in the stellar mass $M_{i}$ bin.

Galaxies are generally categorized into two populations at least out to redshift $z=3-4$, which are known as star-forming galaxies (SFGs) and quiescent galaxies (QGs) \citep{Strateva 2001,Williams 2009,Davidzon 2017}. SFGs follow a relatively tight star formation rate (SFR)$-M_{\rm star}$ relation (the star formation main sequence) up to redshift $z=5-6$ \citep{Noeske 2007,Speagle 2014}. At a given $M_{\rm star}$, SFGs typically have a dispersion of $\sigma_{\rm MS}\sim$0.3 dex in their SFRs \citep{Guo 2013, Speagle 2014}. By contrast, QGs generally have SFRs that are 1-2 dex lower than SFGs. At a given $M_{\rm star}$, QGs are $\sim1$ dex lower in gas fraction compared to SFGs \citep{Spilker 2018,Bezanson 2019}. In this work, we neglect the contribution of gas mass from QGs to the baryonic budget, i.e., the baryonic mass of a QG is assumed to be $M_{\rm baryon, QG}=M_{\rm star, QG}$. In this case, the BMF of QGs has a same form as the SMF.

\begin{figure}
\centering
\includegraphics[width=80mm,angle=0]{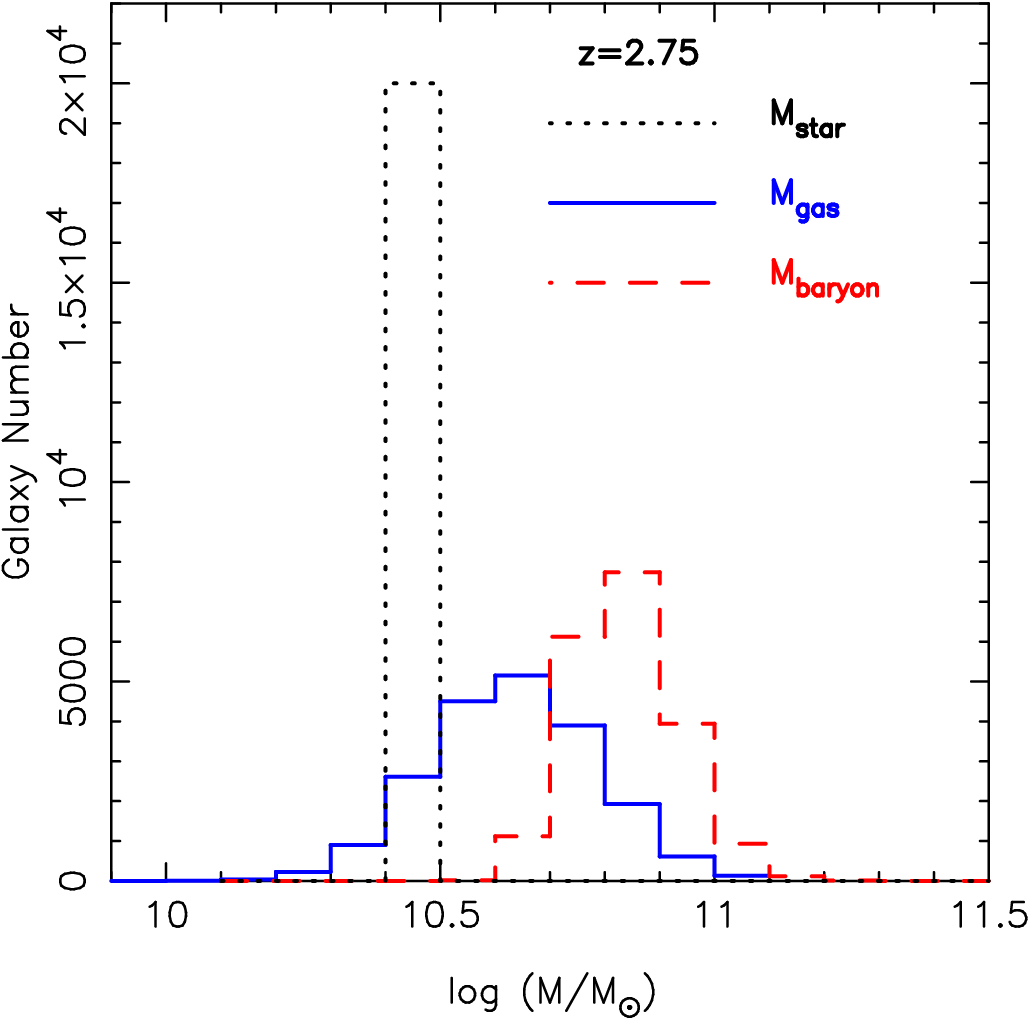}
\caption{An example of our sampling at $z=2.75$ for the star-forming galaxies with log$(M_{\rm star}/M_{\sun})=10.4$. The $M_{\rm star}$, $M_{\rm gas}$ and $M_{\rm baryon}$ distributions are shown in different lines. }\label{fig1}
\end{figure}


For SFGs, the gas content should be considered, i.e., $M_{\rm baryon, SFG}=M_{\rm star, SFG}+M_{\rm gas, SFG}$. The cold gas component of an SFG consists of molecular and atomic hydrogen ($\rm H_{\rm 2}$ and \Hi).  Thanks to the increasing size of galaxies that with CO or far-infrared observations at high redshifts, the properties of molecular gas content of SFGs are extensively investigated in recent years. In \citet{Tacconi 2018}, the authors collected the largest sample to date to investigate the molecular gas content of galaxies in relations to their locations on the main sequence and redshift.  According to \citet{Tacconi 2018}, at a given $M_{\rm star}$, the molecular gas mass of SFGs ($M_{\rm H2}$) has a dispersion of $\sigma=0.52\times \sigma_{\rm MS} \sim 0.15$ dex in the log space when inserting $\sigma_{\rm MS}=0.3$ dex.

The \Hi~content of SFGs can not be studied directly beyond redshift $z=0.4$ due to the present observational limit. Although the evolution of cosmic \Hi~density ($\Omega_{\rm HI}$) at $z<5$ has now been constrained using the observations of damped Ly$\alpha$ systems (DLAs) (see \citealt{Hu 2019}, and reference therein), it is still difficult to use these observations to infer the \Hi~content in the ISM, due to the fact that $\Omega_{\rm HI}$ inferred from DLAs contains the neutral gas that residing in the ISM and circum-galactic medium (CGM). Theoretical studies have suggested that at z>1.75, the majority of $\Omega_{\rm HI}$ is contributed by CGM, rather than ISM \citep{van de Voort 2012, Lagos 2018}. Recently, \citet{Popping 2015} used an indirect technique to infer the evolution of cold ISM from $z=3$ to $z=0.5$, finding that at fixed $M_{\rm star}$, the atomic hydrogen component in the ISM shows no redshift dependence. In this work, we thus use a redshift-independent $M_{\rm star}-M_{\rm HI}$ relation to infer $M_{\rm HI}$ in galaxies.

\begin{figure*}
\centering
\includegraphics[width=160mm,angle=0]{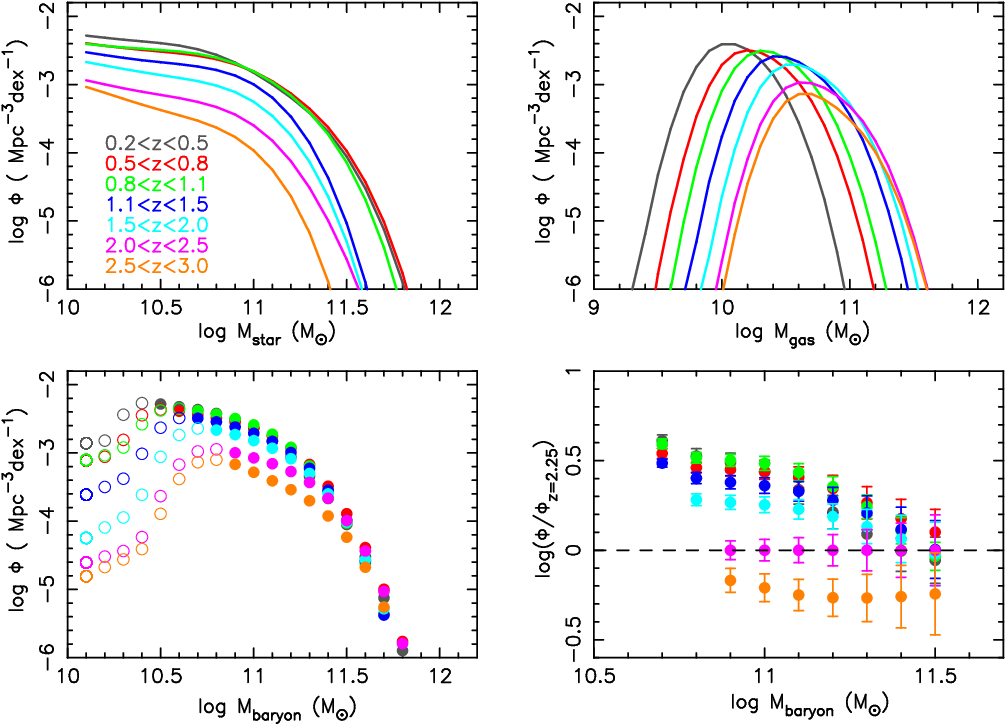}
\caption{Top left: The stellar mass function of \citet{Davidzon 2017}. Top right: The $M_{\rm gas}$ distribution for the log$(M_{\rm star}/M_{\sun})>10.0$ population. Bottom left: The baryonic mass function of galaxies. Solid and open symbols indicates the galaxies with $M_{\rm baryon}\geq M_{\rm limit}$ and $M_{\rm baryon}<M_{\rm limit}$, respectively. Bottom right: Growth of the baryonic mass function relative to the $z=2.25$ redshift bin. We only show the result for the $M_{\rm baryon}>M_{\rm limit}$ galaxies. The errorbars are derived by including the 1$\sigma$ Possian uncertainty of the SMF and $\pm 0.1$ dex uncertainty in $M_{\rm gas}$ estimation. In all panels, galaxies of different redshift bins are indicated by different colors.}\label{fig2}
\end{figure*}

In the local Universe, \citet{Saintonge 2016} found that both $M_{\rm HI}$ and $M_{\rm H2}$ are dependent on the location of SFGs on the MS ($\Delta MS$), in the sense that galaxies above/below the ridge line of the main sequence have elevated/suppressed $M_{\rm HI}$ and $M_{\rm H2}$ compared to mean values. However, the $M_{\rm HI}/M_{\rm H2}$ mass ratio is insensitive to $\Delta MS$. To account for $M_{\rm HI}$, in this work we express the total gas mass $M_{\rm gas}$ as
\begin{equation}
M_{\rm gas}=(1+f)M_{\rm H2},
\end{equation}
where $f=M_{\rm HI}/M_{\rm H2}$. At $z>0$, we assume that $f$ is also insensitive to $\Delta MS$ as found in the local Universe. To derive $M_{\rm gas}$, one needs to quantify $M_{\rm H2}$ and the $f$ factor. In this work, $M_{\rm H2}$ is derived by utilizing the scaling relation of \citet{Tacconi 2018}. On the other hand, $f$ is calculated using the $M_{\rm star}-M_{\rm HI}$ relation of \citet{Pan 2019}, assuming that this relation does not evolve during $z=[0,3]$. As such, in each redshift bin, $f$ can be determined by
\begin{equation}
f=M_{\rm HI}/M_{\rm H2}(\Delta MS=0),
\end{equation}
where $M_{\rm HI}$ at $\Delta MS=0$ is from equation (4) of \citet{Pan 2019}.

We divide galaxies with a mass bin of $\Delta {\rm log}M_{\rm star}=0.1$ dex and use the Monte Carlo method to derive the $M_{\rm baryon}$ distribution of galaxies in each bin. At a given $M_{\rm star}$ bin, we first generated a sample of $N_{\rm gal}$ galaxies and assigned an $M_{\rm gas}$ to each of them. The assigned $M_{\rm gas}$ peaks at the value set by Equation (2), with a dispersion of $\sigma \sim0.15$ dex. By doing so, the $M_{\rm baryon}$ distribution of these $N_{\rm gal}$ galaxies can be determined directly. In Figure~\ref{fig1}, we show an example of the $M_{\rm gas}$ and $M_{\rm baryon}$ distribution of our generated galaxies for the log$(M_{\rm star}/M_{\sun})=10.4$ bin at $z=2.75$. The $M_{\rm baryon}$ distribution of these $N_{\rm gal}$ galaxies is then scaled to match the SMF by multiplying a factor of $\Phi(M_{\rm star,SFG})/N_{\rm gal}$, where $\Phi(M_{\rm star,SFG})$ is the number density of SFGs in that $M_{\rm star}$ bin, which is drawn from the measured SMF. Finally, by summing up the SMF-scaled $M_{\rm baryon}$ distributions for each $M_{\rm star}$ bin, we derive the BMF for SFGs, as illustrated in Equation (1). In this work, we use a sufficiently large galaxy number of $N_{\rm gal}=20000$. Our result is not changed if we choose a larger $N_{\rm gal}$.

The BMF of the global galaxy population is then derived by combing the BMF of SFGs and QGs, where the BMF of QGs has a same form as their SMF as assumed above. In this work we only focus on galaxies with stellar mass greater than log$(M_{\rm star}/M_{\sun})=10.0$, since the gas scaling relation of \citet{Tacconi 2018} is constructed based on galaxies primarily located in this mass regime.

\section{The baryonic mass function}
The SMF used in this work is from \citet{Davidzon 2017}, which is based on the latest dataset of the Cosmic Evolution Survey (COSMOS, \citealt{Scoville 2007}). At each redshift, \citet{Davidzon 2017} used the NUV$-r$ versus $r-J$ diagram to classify galaxies into SFGs and QGs, and measure their SMF separately.  At redshift $z=3$, the SMF is mass completed to log$(M_{\rm star}/M_{\sun})=9.0$, which is sufficiently deep for our study. In the top-left and top-right panel of Figure~\ref{fig2}, we show the SMF and gas mass function (GMF) for the galaxy population of log$(M_{\rm star}/M_{\sun})>10.0$. At $z<1$, it can be seen that the SMF evolves little, while the GMF still has a significant evolution. The bottom-left panel presents our derived BMF. In the low-mass regime, the BMF is highly incomplete because we do not include the galaxy population of log$(M_{\rm star}/M_{\sun})<10.0$. At each redshift, we arbitrarily define a limited mass ($M_{\rm limit}$) below which the effect of incompleteness becomes important, where $M_{\rm limit}=M_{\rm peak}+0.1$ \footnote{$M_{\rm peak}$ is the baryonic mass at which the number density of galaxies reaches the peak value.} . At $M_{\rm baryon}>M_{\rm limit}$, we fit the BMF with a single Schechter function \citep{Schechter 1976}. The $M_{\rm limit}$ and best-fit Schetchter parameters are summarized in Table 1.

Compared to the SMF, a most remarkable feature of the BMF is that at log$(M_{\rm baryon}/M_{\sun})>11.3$, the BMF evolves little since $z\sim 2.2$. Note that we do not account for the uncertainty of SMF in Figure~\ref{fig3}. At very high masses, the BMF has a relatively large uncertainty, which is mainly propagated from the SMF. For the SMF of \citet{Davidzon 2017}, the typical uncertainty is around $\pm0.1$ dex at log$(M_{\rm star}/M_{\sun})=11.0$ and $\pm0.4$ dex at log$(M_{\rm star}/M_{\sun})=11.5$, respectively. In the bottom-right panel, we show the evolution of BMF at $M_{\rm baryon}>M_{\rm limit}$ after accounting for the uncertainty of SMF. We also include a $\pm0.1$ dex uncertainty in the $M_{\rm H2}$ estimation as suggested by \citet{Tacconi 2018}. After accounting for these uncertainties, our finding still holds, suggesting that the baryon assembly is close to completion at the high-mass end ever since the peak of cosmic star formation.

\begin{table*}
\begin{center}
\caption{Schechter parameters of the baryonic mass function}
\begin{tabular}{@{}crrrr}
\tableline\tableline
redshift &mass limit &$\alpha$&$\phi_{\bigstar}$ &$M_{\bigstar}$\\
&log$M_{\rm limit}(h_{70}^{-2}M_{\sun}$)&&$10^{-3}h_{70}^{3}Mpc^{-3}$ &log$M(h_{70}^{-2}M_{\sun})$\\
\tableline
$0.2<z\leq0.5$   &10.5  &$-0.42^{+0.05}_{-0.05}$    &$5.19^{+0.06}_{-0.06}$  &$10.68^{+0.01}_{-0.01}$   \\
$0.5<z\leq0.8$   &10.6  &$-0.77^{+0.05}_{-0.04}$    &$3.45^{+0.14}_{-0.14}$  &$10.88^{+0.02}_{-0.02}$   \\
$0.8<z\leq1.1$   &10.6  &$-0.41^{+0.05}_{-0.04}$    &$4.64^{+0.05}_{-0.05}$  &$10.75^{+0.01}_{-0.01}$ \\
$1.1<z\leq1.5$   &10.7  &$-0.73^{+0.05}_{-0.04}$    &$3.09^{+0.09}_{-0.08}$  &$10.86^{+0.02}_{-0.02}$   \\
$1.5<z\leq2.0$   &10.8  &$-0.60^{+0.05}_{-0.04}$    &$2.38^{+0.05}_{-0.05}$  &$10.85^{+0.01}_{-0.01}$  \\
$2.0<z\leq2.5$   &10.9  &$-0.59^{+0.09}_{-0.08}$    &$1.13^{+0.04}_{-0.04}$  &$10.93^{+0.03}_{-0.03}$  \\
$2.5<z\leq3.0$   &10.9  &$-1.19^{+0.09}_{-0.09}$    &$0.55^{+0.06}_{-0.06}$  &$11.04^{+0.05}_{-0.05}$  \\
\tableline
\end{tabular}
\end{center}
\end{table*}

\section{The baryon net accretion rate and stellar mass growth rate}
Galaxies continuously accrete baryon (from the inter-galactic medium (IGM), or mergers) from the surrounding environments and convert the cold gas into stars. With the evolution of BMF and SMF, we can make a direct comparison between baryon net accretion rate\footnote{net accretion rate=accretion rate$-$outflow rate} and stellar mass growth rate.

\begin{figure}
\centering
\includegraphics[width=80mm,angle=0]{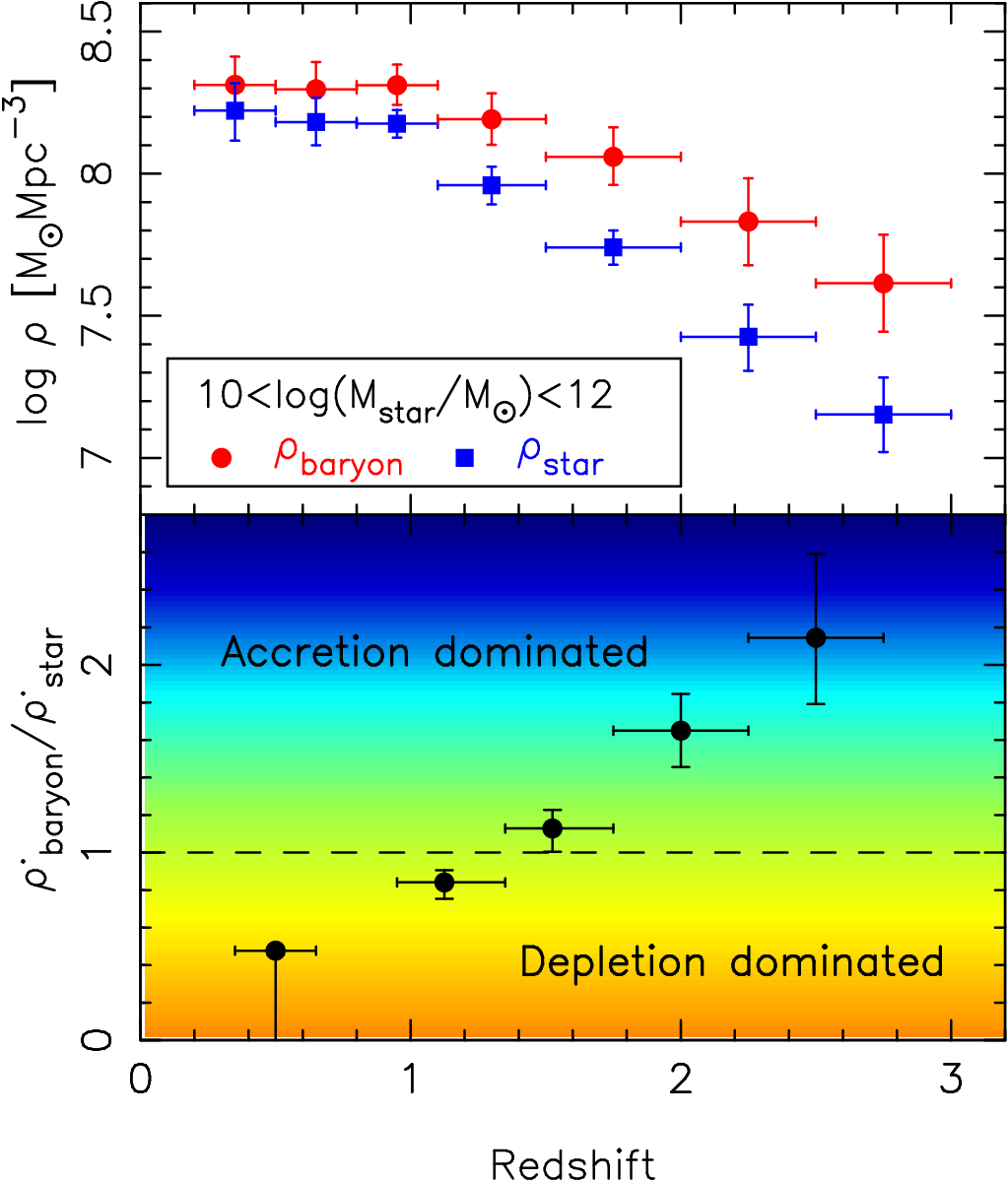}
\caption{Upper panel: The growth of stellar mass density $\rho_{\rm star}$ and baryonic mass density $\rho_{\rm baryon}$ for galaxies with log$(M_{\rm star}/M_{\rm sun})>10.0$. The errorbars are derived by considering the 1$\sigma$ Possian uncertainty in SMF and the uncertainties of BMF shown in Figure~\ref{fig2}. Bottom panel: $\dot{\rho}_{\rm baryon}/\dot{\rho}_{\rm star}$ as a function of redshift. }\label{fig3}
\end{figure}

By convolving the BMF and SMF with $M_{\rm baryon}$ and $M_{\rm star}$ respectively, we derive the stellar mass density $\rho_{\rm star}$ and baryonic mass density $\rho_{\rm baryon}$ for the galaxy population of log$(M_{\rm star}/M_{\sun})>10$.  The upper panel of Figure~\ref{fig3} shows the evolution of $\rho_{\rm star}$ and $\rho_{\rm baryon}$ as a function of redshift.  It can be seen that $\rho_{\rm star}$ and $\rho_{\rm baryon}$ increase by a factor of $\sim 1$ dex ($\times 10$) and $\sim 0.7$ dex ($\times 5$) from $z=3$ to $z=1$, respectively. At $z<1$, both $\rho_{\rm star}$ and $\rho_{\rm baryon}$ shows little evolution. During a specific redshift interval $[z_1,z_2]$, the changing rate of $\rho_{\rm star}$ and $\rho_{\rm baryon}$ can be expressed as:
\begin{equation}
{\dot{\rho}_{\rm star}}=\bigtriangleup \rho_{\rm star}/\Delta t=(\rho_{\rm star,z_1}-\rho_{\rm star,z_2})/\Delta t
\end{equation}
and
\begin{equation}
\dot{\rho}_{\rm baryon}=\bigtriangleup \rho_{\rm baryon}/\Delta t=(\rho_{\rm baryon,z_1}-\rho_{\rm baryon,z_2})/\Delta t,
\end{equation}
where $\bigtriangleup t$ is the time interval between $z_1$ and $z_2$. Hence, at the same redshift interval, $\dot{\rho}_{\rm baryon}$/$\dot{\rho}_{\rm star}$=$\bigtriangleup \rho_{\rm baryon}/\bigtriangleup \rho_{\rm star}$.

The lower panel of Figure~\ref{fig3} shows $\dot{\rho}_{\rm baryon}$/$\dot{\rho}_{\rm star}$ as a function of redshift. At $z\sim 2.5$, $\dot{\rho}_{\rm baryon}$/$\dot{\rho}_{\rm star} \sim 2.0$, indicating that the baryon net accretion rate significantly exceeds the stellar mass growth rate. Assuming that the increasing of $\rho_{\rm baryon}$ and $\rho_{\rm star}$ are primarily due to IGM accretion and stellar mass conversion in galaxies, the global gas density of these galaxies (not for individual) would continuously increase during this epoch. We call this as an ``accretion dominated" phase. $\dot{\rho}_{\rm baryon}$/$\dot{\rho}_{\rm star}$ decreases at later epochs to a level of $\dot{\rho}_{\rm baryon}$/$\dot{\rho}_{\rm star}\sim 1.0$ at $z\sim 1.5$. When $\dot{\rho}_{\rm baryon}$/$\dot{\rho}_{\rm star}<1.0$, the net-accreted baryon is no longer capable in sustaining stellar mass growth, and the gas reservoir of galaxies are being depleted. At $z\sim 0.5$, $\dot{\rho}_{\rm baryon}$/$\dot{\rho}_{\rm star}<0.5$, suggesting that massive galaxies have entered the ``depletion dominated" phase at low redshifts.

It is worthy to emphasize that besides gas accretion and in situ star formation in galaxies, mergers of galaxies with log$(M_{\rm star}/M_{\sun})<10$ will contribute to both ${\rho}_{\rm baryon}$ and ${\rho}_{\rm star}$, hence impacting on the $\dot{\rho}_{\rm baryon}$/$\dot{\rho}_{\rm star}$ ratio \footnote{Mergers between galaxies with log$(M_{\rm star}/M_{\sun})>10$ do not contribute to $\bigtriangleup \rho_{\rm star}$ and $\bigtriangleup \rho_{\rm baryon}$, thus will not impact on the $\dot{\rho}_{\rm baryon}$/$\dot{\rho}_{\rm star}$ ratio.}. Since the low-mass galaxies typically have a higher gas-to-stellar mass ratio than the massive ones, merging of low-mass galaxies will increase the $\dot{\rho}_{\rm baryon}$/$\dot{\rho}_{\rm star}$ ratio.  Assuming that the influence of low-mass mergers is small, the $\dot{\rho}_{\rm baryon}$/$\dot{\rho}_{\rm star}$ in Figure~\ref{fig3} largely reflects the relation between gas net accretion rate and stellar mass conversion rate in galaxies.

\section{Summary and Discussion}
We combine the recently published stellar mass function and the gas scaling relations to explore the evolution of galaxy baryonic mass function of galaxies to redshift $z=3$. We find evidence that at log$(M_{\rm baryon}/M_{\sun})>11.3$, the BMF evolves little since $z\sim 2.2$. By studying the evolution of stellar mass density $\rho_{\rm star}$ and baryonic mass density $\rho_{\rm baryon}$ for the log$(M_{\ast}/M_{\sun})>10$ galaxy population, we find that these galaxies transform from the ``accretion dominated" phase to the ``depletion dominated" phase from high$-z$ to low$-z$, with a transition redshift of $z_{\rm tran}\sim 1.5$.

The robustness of our derived BMF relies on the accuracy of both the SMF measures and $M_{\rm gas}$ estimation. At $z<3$, the SMF measures (both for SFGs and QGs) from different surveys have reached very good agreement for the mass range of log$(M_{\rm star}/M_{\rm \sun})=10.0-11.3$ \citep{Ilbert 2013,Tomczak 2014,Davidzon 2017}. We have also tried the SMF drawn from \cite{Ilbert 2013} and \citet{Tomczak 2014} and found that the change of the BMF is within $\pm 0.1$ dex at log$(M_{\rm baryon}/M_{\sun})<11.3$. At higher masses, the discrepancy increases, which is mainly resulted from cosmic variance. For $M_{\rm gas}$, \citet{Tacconi 2018} stated that the population-average $M_{\rm H2}$ estimation can reach an accuracy of $\pm 0.1$ dex based on their scaling relations. We have considered $\Delta {\rm log}(M_{\rm H2})=0.1~$dex and found that it contributes $\sim0.03-0.07$ dex uncertainty in the BMF at log$(M_{\rm baryon}/M_{\sun})<11.3$. Another source of uncertainty comes from the $M_{\rm HI}$ estimation. In this work we have assumed an unevolved $M_{\rm star}-M_{\rm HI}$ relation at $z<3$. At fixed $M_{\rm star}$, if we allow $M_{\rm HI}$ varies $\pm0.5$ dex since $z=3$ with a simple form of ${\rm log}M_{\rm HI}={\rm log}M_{\rm HI,z=0}+a\times {\rm log}(1+z)$, this will bring in an uncertainty of $\sim0.0-0.12$ dex in the BMF, depending on redshift. So if the $M_{\rm star}-M_{\rm HI}$ relation does not significantly evolve during $z<3$, our BMF should be robust.

The little evolution of BMF at log$(M_{\rm baryon}/M_{\sun})>11.3$ since $z\sim2.2$ implies that the baryon assembly is close to completion in these galaxies since then. How to interpret this phenomenon? At $z<1$, this is another reflection of the little evolution of SMF (see Figure~\ref{fig2}), since the baryon content of massive galaxies has been dominated by stars. At $z>1$, there are some interpretations to this phenomenon under the context of galaxy formation paradigm. One is that the halo masses of these galaxies have exceeded the critical halo mass ($M_{\rm c}\sim 10^{12}M_{\sun}$) to support a stable shock, and the halo gas of these galaxies has been shock-heated to prevent efficient cooling \citep{Dekel 2006}. The deep potential well of massive halos, on the other hand, is capable to prevent strong gas outflows \citep{Tremonti 2004}. The combination of these two effects then naturally result in an extremely low baryon net accretion rate. Alternatively, the low baryon net accretion rate could be resulted from a balance between gas inflows and outflows, without a requirement of a low gas inflow/outflow rate. Some previous works have suggested that high$-z$ massive SFGs could exhibit strong gas outflows due to the strong feedback from super massive black-hole or star formation \citep{Genzel 2014, Yabe 2014}. More future work is needed to better understand this phenomenon.

Interestingly, Figure~\ref{fig2} shows that the number density of galaxies with log$(M_{\rm baryon}/M_{\sun})=11.3$ increases by a factor of 2 ($\sim$ 0.3 dex) from $z\sim 2.7$ to $z\sim 2.2$, implying that the baryon net accretion rate in the progenitors of these galaxies must be considerably high during $z=[2.2,2.7]$. Considering the little evolution of BMF at the massive end since $z\sim 2.2$, some efficient mechanisms are thus required to shutting down the baryon net accretion in massive galaxies within a very short time scale ($\sim 0.5$ Gyr). Under the context of current galaxy formation paradigm, the halo shock heating scenario and feedback from star formation/central black-hole are potentially responsible for doing this job \citep{Dekel 2006,Croton 2006}. Identifying the working mechanism is beyond the scope of this work.

The log$(M_{\rm star}/M_{\sun})>10$ galaxy population transits from the ``accretion dominated" phase to ``depletion dominated" phase at $z_{\rm tran}\sim 1.5$. Interestingly, $z_{\rm tran}$ is similar to the onset redshift of the decline of cosmic star formation rate density (CSFD) (see the review of \citealt{Madau 2014} ). Given that the log$(M_{\rm star}/M_{\sun})>10$ galaxy population contributes to $40-50$\% of the cosmic star formation budget at $z<3.0$, we suggest that the decline of CSFD since $z\sim 1.5$ is closely related to the decline of baryon net accretion rate in galaxies. This is straightforward to interpret since cold gas accretion is necessary for sustaining star formation \citep{Dave 2011,Lilly 2013,Peng 2014}. With the decline of baryon accretion rate, the star formation rate in galaxies will decrease, which then results in star formation cessation in some galaxies, especially at the high-mass end \citep{Peng 2010, Pan 2016, Pan 2017}. The early assembly of baryon content in massive galaxies at $z\sim 2.2$ and the rapid build up of massive QG population since then support this picture \citep{Ilbert 2013,Muzzin 2013, Tomczak 2014,Davidzon 2017}. In summary, our findings support the idea that the decline of cosmic star formation rate density since $z\sim1.5$ is mainly resulted from the decline of baryon net accretion rate and star formation quenching in galaxies.

\acknowledgments
We thank the anonymous referee for constructive suggestions that help improving the clarity of the manuscript. This work was partially supported by the National Key Research and Development Program (``973" program) of China (No.2015CB857004, 2016YFA0400702, 2017YFA0402600 and 2017YFA0402703), the National Natural Science Foundation of China (NSFC, Nos. 11773001, 11703092, 11320101002, 11421303, 11433005, 11773076 and 11721303), and the Natural Science Foundation of Jiangsu Province (No.BK20161097).

\end{document}